\documentclass[lettersize,journal]{IEEEtran}
\usepackage{amsmath,amsfonts}
\usepackage{algorithmic}
\usepackage{array}
\usepackage{cite}
\usepackage[caption=false,font=normalsize,labelfont=sf,textfont=sf]{subfig}
\usepackage{svg}
\usepackage{textcomp}
\usepackage{url}
\usepackage{verbatim}
\usepackage{multirow}

\usepackage{graphicx}
\graphicspath{ {figures/} }

\usepackage{dblfloatfix}

\usepackage{color}
\usepackage{xcolor}
\usepackage{listings}
\definecolor{light-gray}{gray}{0.95} 

\def\BibTeX{{\rm B\kern-.05em{\sc i\kern-.025em b}\kern-.08em
    T\kern-.1667em\lower.7ex\hbox{E}\kern-.125emX}}
\usepackage{balance}
\begin{document}
\title{Toward Building a Semantic Network Inventory for Model-Driven Telemetry}
\author{Ignacio D. Martinez-Casanueva, Daniel Gonzalez-Sanchez, Luis Bellido, David Fernandez, and Diego R. Lopez
\thanks{I. D. Martinez-Casanueva (email: i.dominguezm@alumnos.upm.es), D. Gonzalez-Sanchez, L. Bellido, D. Fernandez are with ETSI Telecomunicacion, Universidad Politecnica de Madrid, Spain.
D. R. Lopez is with Telefonica I+D, Spain.}
\thanks{This is the authors’ accepted version of the article. The final version published by IEEE is I. D. Martínez-Casanueva, D. González-Sanchez, L. Bellido, D. Fernández and D. R. López, "Toward Building a Semantic Network Inventory for Model-Driven Telemetry," in IEEE Communications Magazine, vol. 61, no. 3, pp. 60-66, March 2023, doi: 10.1109/MCOM.001.2200222.}
\thanks{©2023 IEEE. Personal use of this material is permitted. Permission from IEEE must be obtained for all other uses, in any current or future media, including reprinting/republishing this material for advertising or promotional purposes, creating new or redistribution to servers or lists, or reuse of any copyrighted component of this work in other work.}}

\maketitle

\begin{abstract}
Network telemetry based on data models is expected to become the standard mechanism for collecting operational data from network devices efficiently. But the wide variety of standard and proprietary data models along with the different implementations of telemetry protocols offered by network vendors, become a barrier when monitoring heterogeneous network infrastructures. To facilitate the integration and sharing of context information related to model-driven telemetry, this work proposes a semantic network inventory that integrates new information models specifically developed to capture context information in a vendor-agnostic fashion using current standards defined for context management. To automate the integration of this context information within the network inventory, a reference architecture is designed. Finally, a prototype of the solution is implemented and validated through a case study that illustrates how the network inventory can ease the operation of model-driven telemetry in multi-vendor networks.  

\end{abstract}


\section{Introduction}
Network infrastructures encompass a wide variety of monitoring data sources. To access the monitoring data, each data source entails different techniques, which can be classified in two categories: pull or push methods. Similarly, data sources may provide data in different formats, ranging from raw to structured formats based on standard data models. Therefore, to cope with such heterogeneity, it is crucial to manage the metadata that captures the characteristics of the monitoring data sources.

Among the different monitoring techniques proposed for network infrastructures, model-driven telemetry (MDT) is drawing significant interest \cite{Claise_book}. MDT builds upon streaming monitoring data from network devices that are modeled by means of the YANG language. YANG enables defining formal data models for managing specific features of a network device such as network interfaces or BGP sessions. MDT provides efficient mechanisms for subscribing to specific YANG data in a model and for pushing the data, either periodically or upon changes, to data collectors. In this sense, MDT enables near real-time monitoring of a network infrastructure, for example, by dynamically detecting a shutdown of an interface, or by identifying network anomalies based on BGP information \cite{mdt_bgp}. For these reasons, YANG data models, and consequently MDT, are considered key enablers for network monitoring automation \cite{Pastor2021}.

Since the inception of the YANG language, the industry has experienced a massive growth of YANG data models developed by network vendors (e.g., Cisco), standards developing organizations (e.g., ETSI), open-source projects (e.g., OpenDaylight), and consortia (e.g., OpenConfig) \cite{yang_stats}. Moreover, many of these data models may augment or deviate others to define new features or remove existing features depending on the device implementation. Table \ref{tab1} captures the total number of standard and proprietary YANG models implemented by three tested network devices from different vendors. The results show the big difference between the number of proprietary models versus the number of standard models, but also that the tested devices only have 11 standard models in common. Therefore, dealing with this myriad of data models, especially when it comes to proprietary models, presents a challenge for network service providers, which need to operate large multi-vendor network infrastructures.

\begin{table}[t]
\fontsize{7.6}{10}\selectfont
\begin{center}
\caption{YANG data models supported by different network devices}
\label{tab1}
\begin{tabular}{|c|c|c|c|c|}
\hline
\begin{tabular}[c]{@{}c@{}}Network device\end{tabular}     & \begin{tabular}[c]{@{}c@{}}Standard\\ models\end{tabular} & \begin{tabular}[c]{@{}c@{}}Common\\ standard\\ models\end{tabular} & \begin{tabular}[c]{@{}c@{}}Proprietary\\ models\end{tabular} & \begin{tabular}[c]{@{}c@{}}Proprietary\\ deviations\end{tabular} \\ \hline
\begin{tabular}[c]{@{}c@{}}Cisco IOS-XRv 9000\end{tabular} & 104                                                       & \multirow{3}{*}{11}                                                & 817                                                          & 23                                                               \\ \cline{1-2} \cline{4-5} 
\begin{tabular}[c]{@{}c@{}}Huawei NE40E-X8\end{tabular}    & 103                                                       &                                                                    & 658                                                          & 38                                                               \\ \cline{1-2} \cline{4-5} 
\begin{tabular}[c]{@{}c@{}}Nokia 7750 SR\end{tabular}      & 84                                                        &                                                                    & 111                                                          & 40                                                               \\ \hline
\end{tabular}
\end{center}
\end{table}

When managing MDT-enabled devices, network engineers first need to find what YANG data are available in each device and how these data can be streamed through an MDT network protocol. These MDT metadata are usually scattered across isolated domains of information, which range from curated registries of YANG data models to manifests containing details on the supported network protocols. The current approach followed in the industry requires network engineers to figure out the location of these MDT metadata, which will vary depending on the vendor. This approach is slow and costly since it relies on the network engineer's individual knowledge to manually search and collect MDT metadata.

In this paper, we propose a semantic network inventory to gather and share MDT metadata collected from different isolated domains. The inventory builds on the ETSI Industry Specification Group (ISG) Context Information Management (CIM) standard, which facilitates the integration of context information (e.g., metadata) in the form of a property graph. The contributions of the semantic network inventory are twofold: (i) the definition of novel ETSI CIM information models that represent MDT context information from various domains; (ii) a reference architecture that leverages ETSI CIM mechanisms to automate the integration of MDT context information collected from the different domains, as well as to expose such information through a centralized and standardized interface.

The remainder of this paper is structured as follows. Section II enumerates existing solutions to manage YANG data models. Then, the ETSI CIM standard and related works are presented. Section III introduces new information models that capture context information related to MDT. Section IV describes a reference architecture that specifies context exchanging mechanisms for building a network inventory. Section V validates the proposal with a use case that measures performance indicators of network interfaces in devices from different vendors. Finally, Section VI draws conclusions and proposes some ideas for future work.

\section{Related Work}

\subsection{YANG data models and YANG modules}
YANG data models are defined in YANG modules, which are files that contain specifications of the structure of the data. A data model can be composed of multiple modules, which in turn, may be split into submodules. In this regard, there are different approaches for searching and managing YANG modules, which can be sorted into three types: repository, catalog, and explorer.

A \textit{repository} represents a collection of files of YANG modules that are publicly available. Network vendors provide repositories that arrange YANG modules based on how these are implemented for each network device model. Other repositories gather YANG modules that are specifically developed by a consortium such as OpenConfig, whereas initiatives like YangModels \cite{yang_models} aim to concentrate YANG modules that span across different vendors, consortia, or open-source projects. In summary, there is a plethora of repositories with different purposes, making it difficult to manage the existing YANG modules.

A \textit{catalog} can be defined as a solution that extracts the metadata from YANG modules – typically stored in a repository – and exposes these metadata through a comprehensive search system. The YANG Catalog \cite{yang_catalog} is a project driven by the IETF to implement a centralized catalog that operators can leverage to search for YANG modules. Modules can be filtered based on metadata such as the name of the organization that developed the module. For these reasons, catalogs are the most complete solution for finding YANG modules, however, catalogs are managed as silos of information that are unaware of how YANG modules are actually implemented by devices in a network.

An \textit{explorer} is identified as a tool that can connect to devices running in a network and learn the YANG modules that are supported. For example, Cisco YANG Suite \cite{yang_suite} implements an explorer that provides a graphical interface for browsing YANG modules supported by the device. Additionally, YANG Suite allows for sending RPCs to edit the configuration and get monitoring data of a device in an interactive fashion. Therefore, explorers are convenient tools that facilitate learning and interacting with the YANG modules supported by devices in a network. Nevertheless, explorers are built as interactive tools for ad-hoc testing, which prevents linking the information related to devices with the information provided by other elements such as catalogs.

\subsection{ETSI CIM standard}
The ETSI CIM standard considers the exchange of context information across systems to be a crucial enabler for smart applications. The standard allows these applications to collect information from different sources, integrate the information, and eventually, generate derivative information or make decisions. To this end, CIM specifies a protocol named NGSI-LD for exchanging context information between smart services. The NGSI-LD protocol comprises two main innovations: the NGSI-LD information model and the NGSI-LD API.

The NGSI-LD information model \cite{cim_006} derives from the labeled property graph model where entities are connected through relationships. Both entities and relationship are conceived as objects whose characteristics are represented using properties. NGSI-LD defines these foundational classes – entity, property, relationship – in the so-called NGSI-LD meta-model, which represents the formal basis for building property graphs using standards of the Semantic Web. This allows context information in a graph to refer  public ontologies which can be leveraged to apply semantic reasoning. Furthermore, the NGSI-LD model extends the expressiveness of the labeled property graph model by adding support for second-level logic, which enables defining properties-of-properties, relationships-of-properties, and relationships-of-relationships.

The NGSI-LD API \cite{cim_009} defines a RESTful-based API as the means for exchanging context information. ETSI CIM does not specify any reference architecture, though the standard introduces multiple architectural roles that exchange information through the Context Broker by using the NGSI-LD API. The Context Producer and Context Source roles register, provide, and update the context information they can offer. Meanwhile, the Context Consumer role discovers relevant context information and subscribes for notifications. 

Thus far, IoT has been the main domain for the application of the CIM framework. López-Morales et al. \cite{Lopez-Morales2020} propose a monitoring platform for the agriculture sector. In this work, the authors demonstrate how IoT sensors can be integrated as Context Producers using the standard. An intermediate component called IoT Agent is responsible for collecting data from sensors, adapting the data to the NGSI-LD format, and sending the data to the Context Broker. Similarly, within the scope of smart cities, Jeong et al. \cite{Jeong2020} present a platform that integrates heterogeneous data sources by implementing adaptors that ingest data from each source and convert the data into the NGSI-LD format. These contributions have served as reference for the implementation of Context Producers within the scope of network monitoring, which is a topic that has not been addressed with ETSI CIM thus far.

\begin{figure*}[bp]
\centering
\includegraphics[width=4.6in]{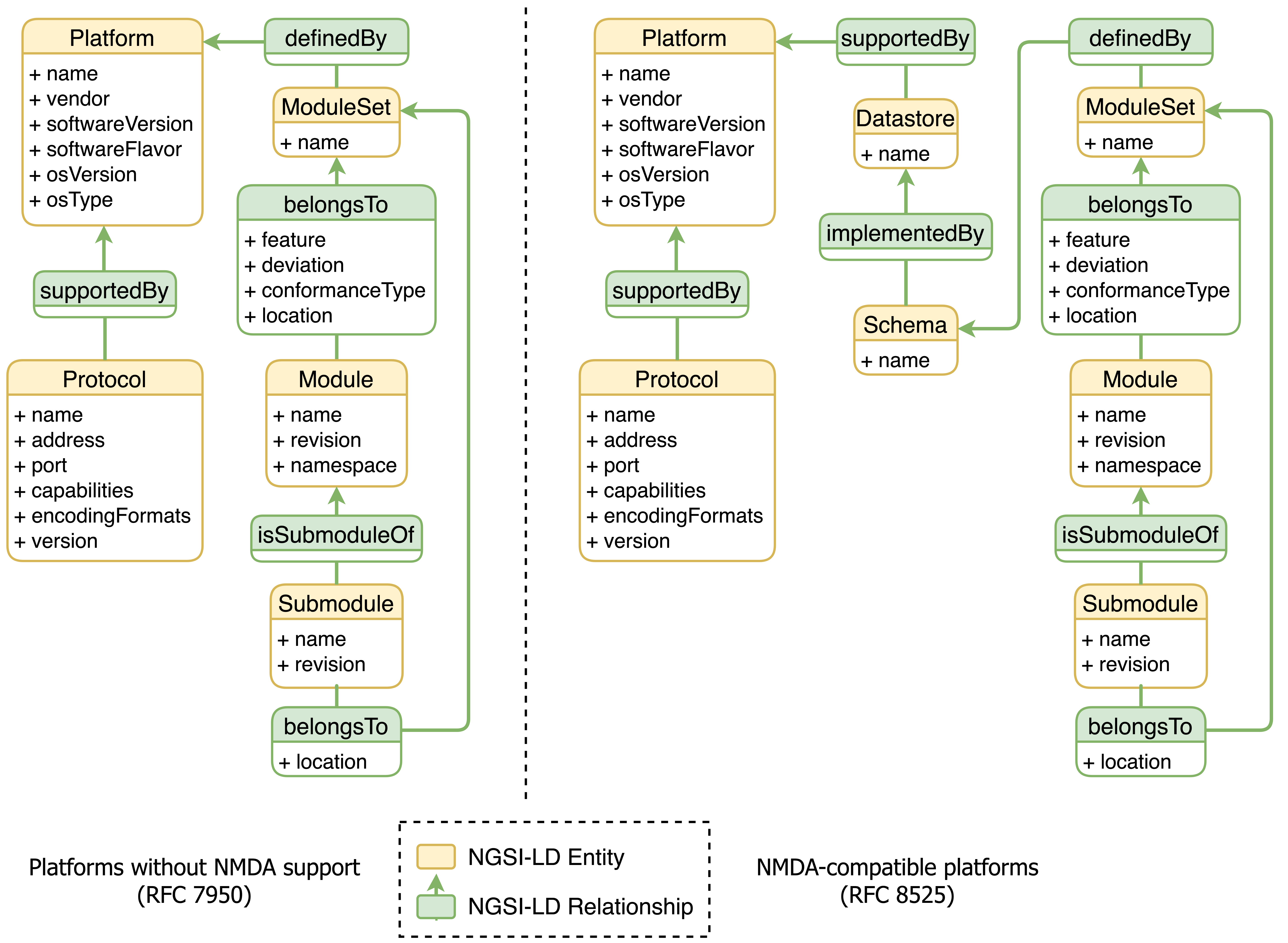}
\caption{NGSI-LD information models for NMDA and non-NMDA platforms.}
\label{fig:platform_domain}
\end{figure*}

\section{NGSI-LD Information Models for MDT-Based Monitoring}

The ETSI CIM standard defines context information as any relevant information about entities, their properties, and their relationships with other entities, thus composing a graph. Entities may be representations of real-world objects, such as physical servers or network devices, but may also describe abstract concepts such as a virtualized network function (VNF) within the scope of the network function virtualization (NFV) paradigm. In this way, we could regard the context information of a system as being all the relevant operational aspects that are required for it to work as intended.

When it comes to context information related to MDT, we identify two relevant domains of information. First, the \textit{Platform Domain}, which contains information regarding the implementation of YANG modules and MDT protocols in network platforms – hereafter the term ``platform'' will be used instead of ``device'' in order to encompass other network elements such as software defined networking (SDN) controllers. Second, the \textit{Catalog Domain}, which includes metadata about the YANG modules existing in external YANG catalogs. To enable the integration of information from these two domains, we propose novel NGSI-LD information models that have concepts (i.e., entities) in common. In our case, the Platform Domain and the Catalog Domain share the YANG module and YANG submodule entities.

To design new NGSI-LD information models, ETSI CIM recommends referencing public ontologies in order to leverage the Semantic Web standards. However, in the use case addressed in this work, we did not find ontologies that cover the implementation details of YANG modules in a network platform neither additional information regarding YANG modules themselves. Notwithstanding, we identified YANG modules from which to derive the proposed NGSI-LD information models. Precisely, the standardized YANG Library module \cite{rfc8525} binds to the Platform Domain, whereas the YANG Catalog project defines another YANG module \cite{clacla-netmod-model-catalog-03} that serves as reference for the Catalog Domain.

\subsection{Platform Domain}

The details on the implementation of YANG modules and submodules by network platforms are best defined by the YANG Library module, which adds support for network platforms compatible with the Network Management Datastore Architecture (NMDA). The NMDA introduces a new architecture whereby a platform contains separate datastores for running configuration and current operational data, but may also include more datastores such as startup or candidate configurations.

Our proposal for the Platform Domain is to design two NGSI-LD information models that derive from the YANG Library module: a model for NMDA-compliant platforms and another model for platforms that do not support NMDA. Fig. \ref{fig:platform_domain} illustrates the two NGSI-LD information models that result from analyzing the semantics of the YANG Library module and mapping YANG to NGSI-LD. Both information models can be split into two groups: (i) platform and the available network management protocols; (ii) implementation of YANG modules and submodules in the platform.

The first group is common to both information models. The top entity is {\it Platform}, which represents a network platform. Secondly, MDT-enabled platforms expose services that allow for streaming telemetry by using network management protocols such as Network Configuration Protocol (NETCONF) and gRPC Network Management Interface (gNMI). The information model records the configuration of the services implemented by a network platform with the {\it Protocol} entity type. The {\it Protocol} type contains properties such as the IP address and port through which the service is reachable, the capabilities supported by the protocol, and the supported encoding formats (e.g., Protobuf).

The second group will vary for each information model depending on whether the network platform is compatible with NMDA or not. In the case of NMDA-compatible platforms, the different available datastores are represented by the {\it Datastore} entity that, in turn, contains a datastore schema represented by the {\it Schema} entity. A datastore schema is the union of {\it ModuleSet} entities which consist of collections of {\it Module} and {\it Submodule} entities. The details regarding the implementation of each {\it Module} and {\it Submodule} in a {\it ModuleSet} are described with the {\it belongsTo} relationship. For example, the {\it conformanceType} property determines whether a YANG module is implemented or only imported by other modules (i.e., not protocol-accessible in the platform). Conversely, in the case of platforms that are not compatible with NMDA, the concept of separate datastores is not supported. Therefore, the resulting information model does not include the {\it Datastore} and {\it Schema} entities, but a single {\it ModuleSet} entity that is implemented by the platform.

\subsection{Catalog Domain}

As a result of the NGSI-LD information models designed for the Platform Domain, the generated context information would result into entities of YANG modules and submodules that only contain mandatory identifiers: name, revision, and namespace. But frequently, consumers of this context information will need further details than the identifiers of YANG modules implemented by platforms in a network. Context information of a YANG module such as an RFC document, dependencies with other modules, or even the location of the code, will help at understanding the meaning of the data. In this regard, the YANG Catalog initiative stores such context information and internally models the information as per the YANG Catalog module.

The YANG Catalog module comprises two main sub-trees: the Module sub-tree and the Vendor sub-tree. We will focus on the former as the latter is related with implementation details by network vendors, which is precisely what the NGSI-LD information models for the Platform Domain aim to provide. Therefore, the Module sub-tree will serve as reference to build an NGSI-LD information model for the Catalog Domain. 

\begin{figure}[!t]
\centering
\includegraphics[width=2.3in]{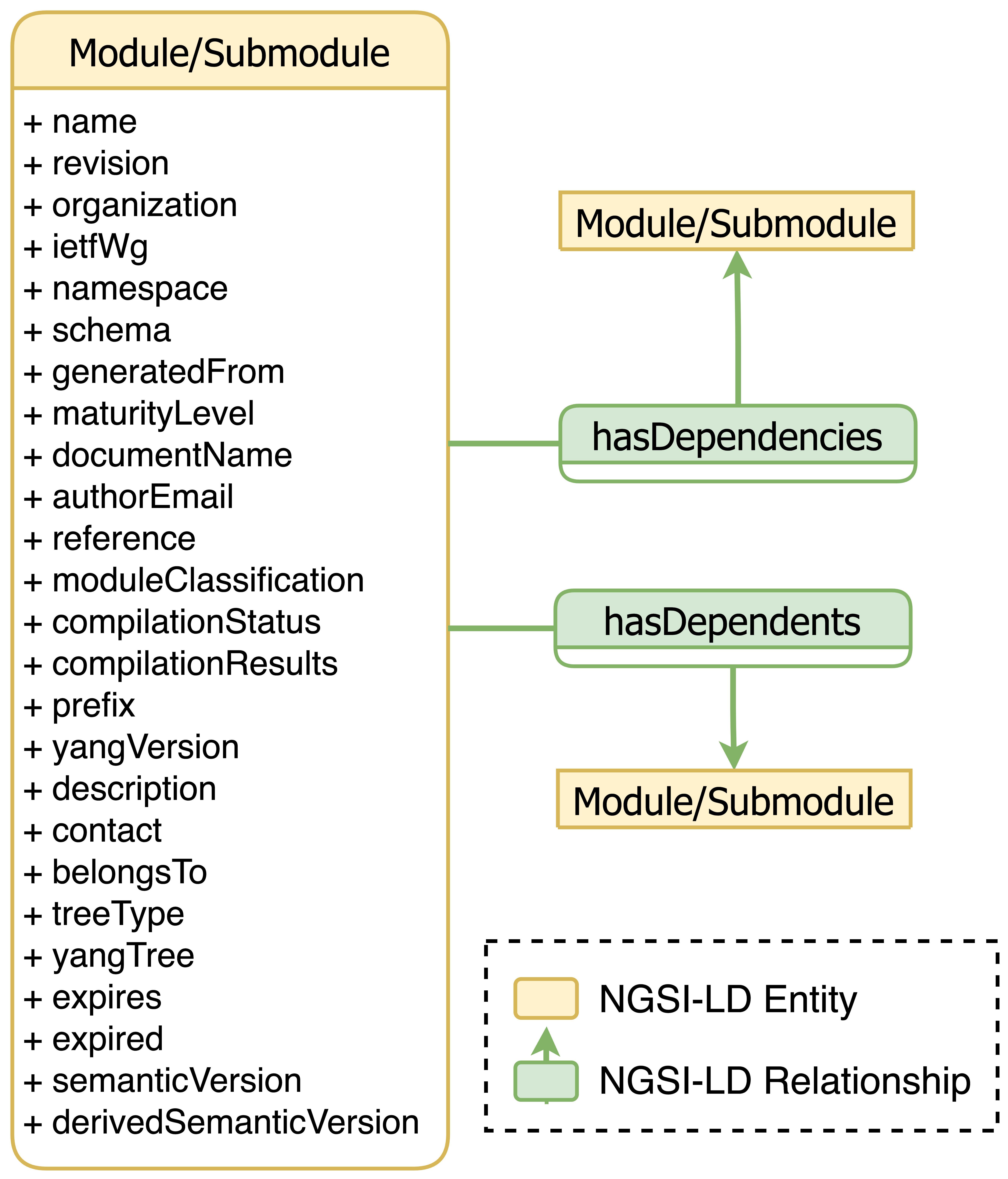}
\caption{NGSI-LD information model for the Catalog Domain.}
\label{fig:catalog_domain}
\end{figure}

After analyzing the semantics of the Module sub-tree, the resulting NGSI-LD information model is depicted in Fig. \ref{fig:catalog_domain}. The information model enriches the {\it Module} and {\it Submodule} entity types that were previously introduced for the Platform Domain. These entity types share the same structure, which contains a deep list of properties representing portions of YANG metadata. For instance, the {\it schema} property provides a URL to the YANG code in a public repository; the {\it treeType} property describes whether the model follows a particular structure such as NMDA or OpenConfig; the {\it semanticVersion} property conveys the version of the model as standardized by OpenConfig. Lastly, the {\it Module} or {\it Submodule} entity may include two different relationships: {\it hasDependencies} and {\it hasDependents} relationships, which convey modules or submodules that import or include others and vice-versa.

\section{Network Inventory Architecture}

The Network Inventory implements the pieces that enable integrating context information associated with MDT. The purpose of the Network Inventory is to enable Network Operations (NetOps) engineers to discover the available YANG data (i.e., \textit{what}) and the network management protocols that can stream such data (i.e., \textit{how}) from a given platform. The NetOps engineer is a new role that, by applying DevOps techniques, can develop applications that automate the management of network infrastructures. In this sense, based on the insights collected from the Network Inventory, NetOps engineers can autonomously develop network monitoring applications supported by MDT, without having to interact manually with external resources or other engineers in the team.

\begin{figure}[!t]
\centering
\includegraphics[width=3.15in]{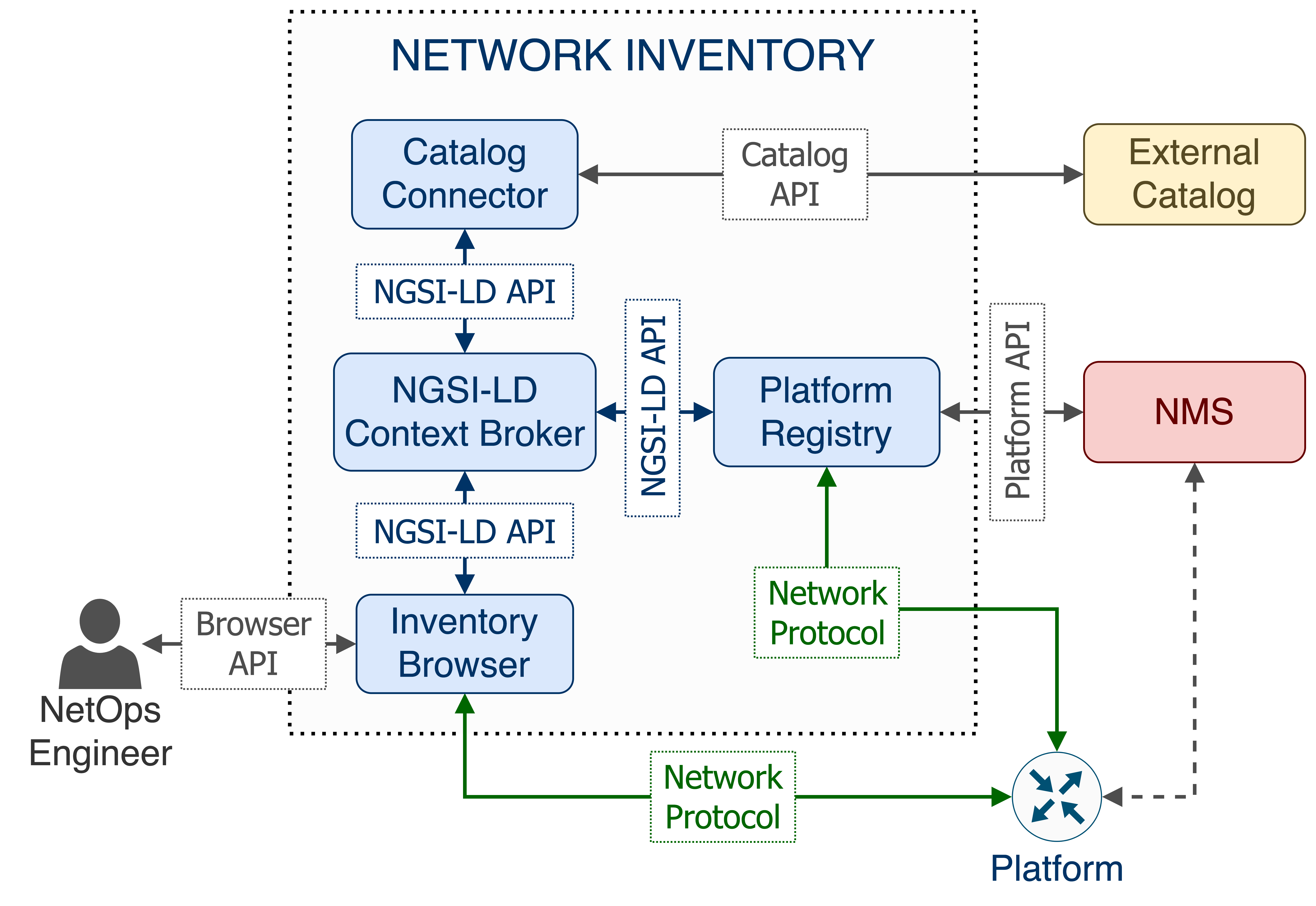}
\caption{Reference architecture of the Network Inventory.}
\label{fig:inventory}
\end{figure}

Fig. \ref{fig:inventory} depicts the envisioned architecture for the Network Inventory which is composed of the following main building blocks: NGSI-LD Context Broker, Platform Registry, Catalog Connector, and Inventory Browser. The Context Broker sits at the core of the architecture, storing context information and exposing a unified interface for context searching through the NGSI-LD API. 

The Platform Registry listens for notifications of platforms that are registered in the network by a Network Management System (NMS). Once a platform gets registered, the Platform Registry discovers the capabilities of the target platform and writes the resulting context information, which follows the Platform Domain model, in the Context Broker. 

The Catalog Connector integrates external catalogs such as the IETF YANG Catalog into the Network Inventory. These types of catalogs serve as great sources of metadata related with YANG modules and submodules. In this sense, the Catalog Connector represents the component that produces context information modeled according to the Catalog Domain model.

The Inventory Browser acts as the frontend of the Network Inventory providing APIs and a GUI for NetOps engineers to explore the capabilities of platforms registered in the Network Inventory. In this sense, the Inventory Browser consumes context information stored in the NGSI-LD Context Broker and presents it to NetOps engineers in an interactive fashion. Additionally, the Inventory Browser implements clients for different network management protocols, thus, allowing operators for interacting with platforms registered in the Network Inventory. The intended behavior is similar to that implemented by the aforementioned network explorers.

The following subsections introduce the two main features implemented in the Network Inventory.  Each subsection describes in detail how context information is exchanged among the different building blocks.

\subsection{Platform registration}

Automatic collection of capabilities is one of the main features that the Network Inventory implements during the registration of network platforms. Once a new platform is added to the inventory, the model-based capabilities of the platform  (i.e., the implemented YANG Modules, and potentially, the available datastores) are added as context information in the NGSI-LD Context Broker. Thus, the Platform Registry implements an “on-demand” Context Producer.

The registration process starts when the Platform Registry receives a registration event. A registration event may be generated in two possible ways: a) the Platform Registry exposes an API through which the NMS requests a platform registration; b) the Platform Registry listens to events from the NMS that trigger a platform registration. In both approaches, the registration event must include the minimum required information for the Network Inventory to connect to the network platform, namely, IP address and port of the network management protocol(s), and credentials if needed.

Based on the connection details, the Platform Registry selects a client that implements the network management protocol(s) supported by the network platform. The chosen client sends the corresponding capability discovery request to the platform. The returned message includes further information about the protocol such as capabilities and encoding formats, along with the supported YANG modules. 

Since the proposed NGSI-LD information model for the Platform Domain derives from the YANG Library module, the Platform Registry first certifies that this YANG module is implemented in the platform by inspecting the contents of the capability discovery reply. In such a case, the Platform Registry retrieves data available from this module and translates the results into to the NGSI-LD information model accordingly (i.e., populating the graph with datastore information or only with module-set information when NMDA is not supported).

Conversely, for those platforms that do not support the YANG Library module, the Platform Registry implements a fallback mechanism to support backwards compatibility. This mechanism evaluates the list of YANG modules retrieved from the capability discovery request that was sent to the platform. However, compared to fetching data from the YANG Library module, this mechanism cannot obtain all context information such as which YANG modules are only imported by the platform (i.e., the \textit{conformanceType} property). In this sense, the implementation of the fallback mechanism must cope with the particularities of each network protocol in order to determine which \textit{Module} entities are linked with a \textit{Platform} entity in the NGSI-LD information model.

\subsection{YANG metadata collection}

The second feature supported by the Network Inventory is the exploration of the metadata of YANG modules. Once a platform has been successfully registered, the context information stored in the Context Broker shows a \textit{Platform} entity implementing \textit{Module} entities, but further information about the \textit{Module} entities needs to be obtained.

To provide additional context information for \textit{Module} and \textit{Submodule} entities, the Catalog Connector component is envisioned. This component collects context information concerning \textit{Module} and \textit{Submodule} entities by pulling metadata from an External Catalog, and then stores it in the Context Broker. Hence, the Catalog Connector is designed as a “scheduled” Context Producer. This design is chosen because External Catalogs are commonly managed by an external party (e.g., IETF) and they store metadata that change with low frequency.

The Catalog Connector periodically retrieves all metadata from an External Catalog. In this process, the Catalog Connector performs a mapping between the Catalog Domain model – shown in section 3 – and the data model of the External Catalog. As a result, the metadata collected from the External Catalog are translated into context information of \textit{Module} and \textit{Submodule} entities that are updated in the Context Broker.

\section{Case Study: Measuring Network Interface Performance}

This section shows how the proposed Network Inventory facilitates measuring network interface KPIs such as throughput or packet loss through MDT mechanisms. First, a prototype aligned with the Network Inventory’s architecture is introduced. Second, we compare how the traditional method and our inventory-based method address this use case.

\subsection{Validation prototype}

\begin{figure}[!t]
\centering
\includegraphics[width=3.15in]{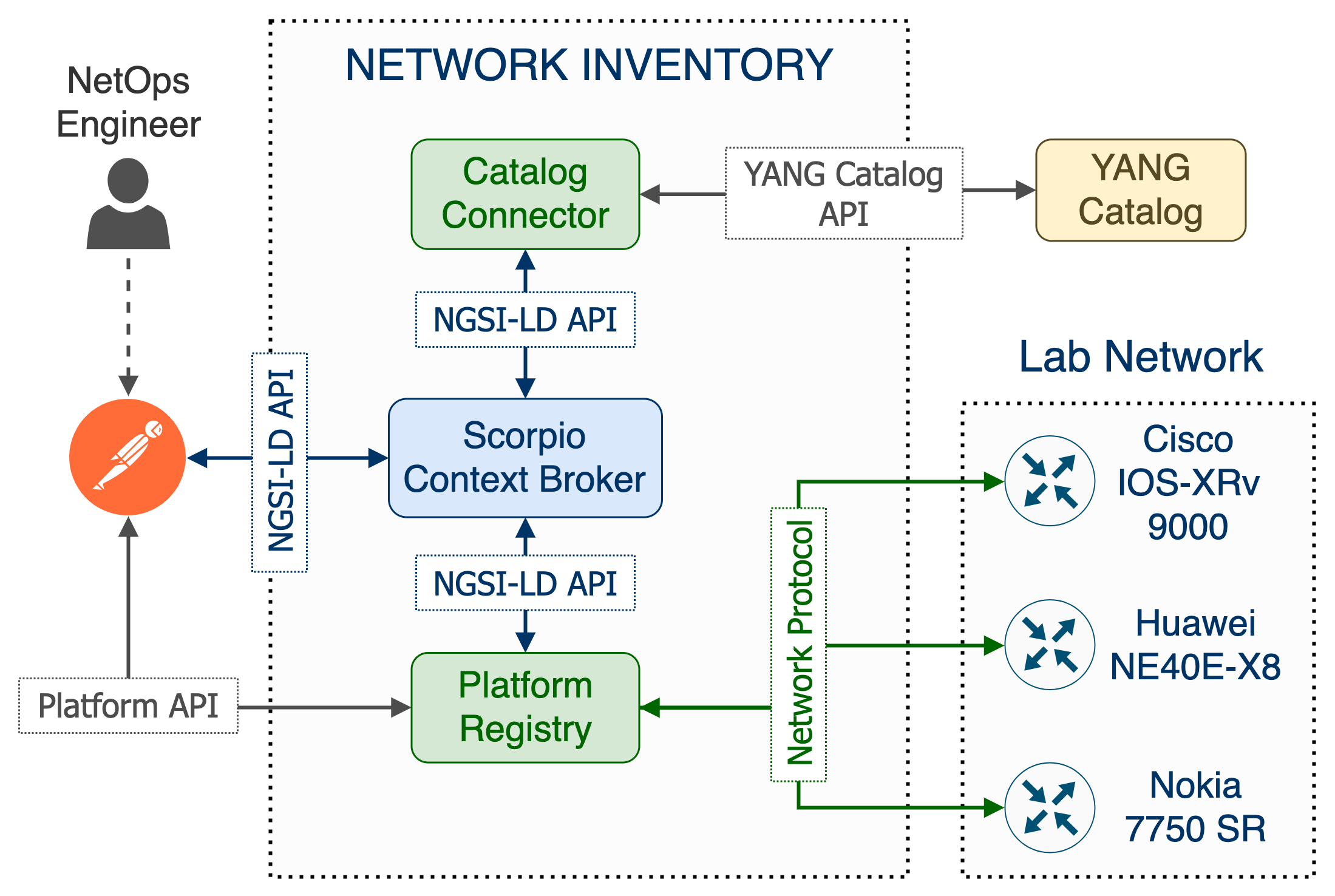}
\caption{Network Inventory prototype.}
\label{fig:prototype}
\end{figure}

A prototype of the Network Inventory is implemented as depicted in Fig. \ref{fig:prototype}. The code that was developed is publicly available on GitHub \cite{network-inventory}. The components comprising the Network Inventory are containerized using Docker and deployed using docker-compose. Scorpio \cite{scorpio} is selected as the implementation of the NGSI-LD Context Broker. The Catalog Connector component is a Python application that implements clients for the NGSI-LD API and YANG Catalog API, since YANG Catalog is chosen as the example External Catalog for the validation. The Platform Registry component is a Python application that exposes a REST API to allow the registration of network platforms in the Network Inventory, acting as a client for the NGSI-LD API. Additionally, the Platform Registry implements clients for both the NETCONF protocol and the gNMI protocol that will be used to discover capabilities of registered network platforms. 

To validate the prototype, an experimental scenario has been provisioned in Telefonica's premises. The scenario runs a heterogeneous network comprising devices from various network vendors, namely, Cisco IOS-XRv 9000, Huawei NE40E-X8, and Nokia 7750 SR. In this context, these devices are MDT-enabled network platforms that will be registered in the Network Inventory during the validation.

Note that the prototype does not include the Inventory Browser component. Ideally, this component would provide a GUI to query the Context Broker and link NGSI-LD entity identifiers on behalf of the engineer. For convenience, this prototype leverages the Postman API tool to facilitate the interactions with the Network Inventory.

\subsection{Method comparison}

\begin{figure}[!b]
\centering
\includegraphics[width=3.48in]{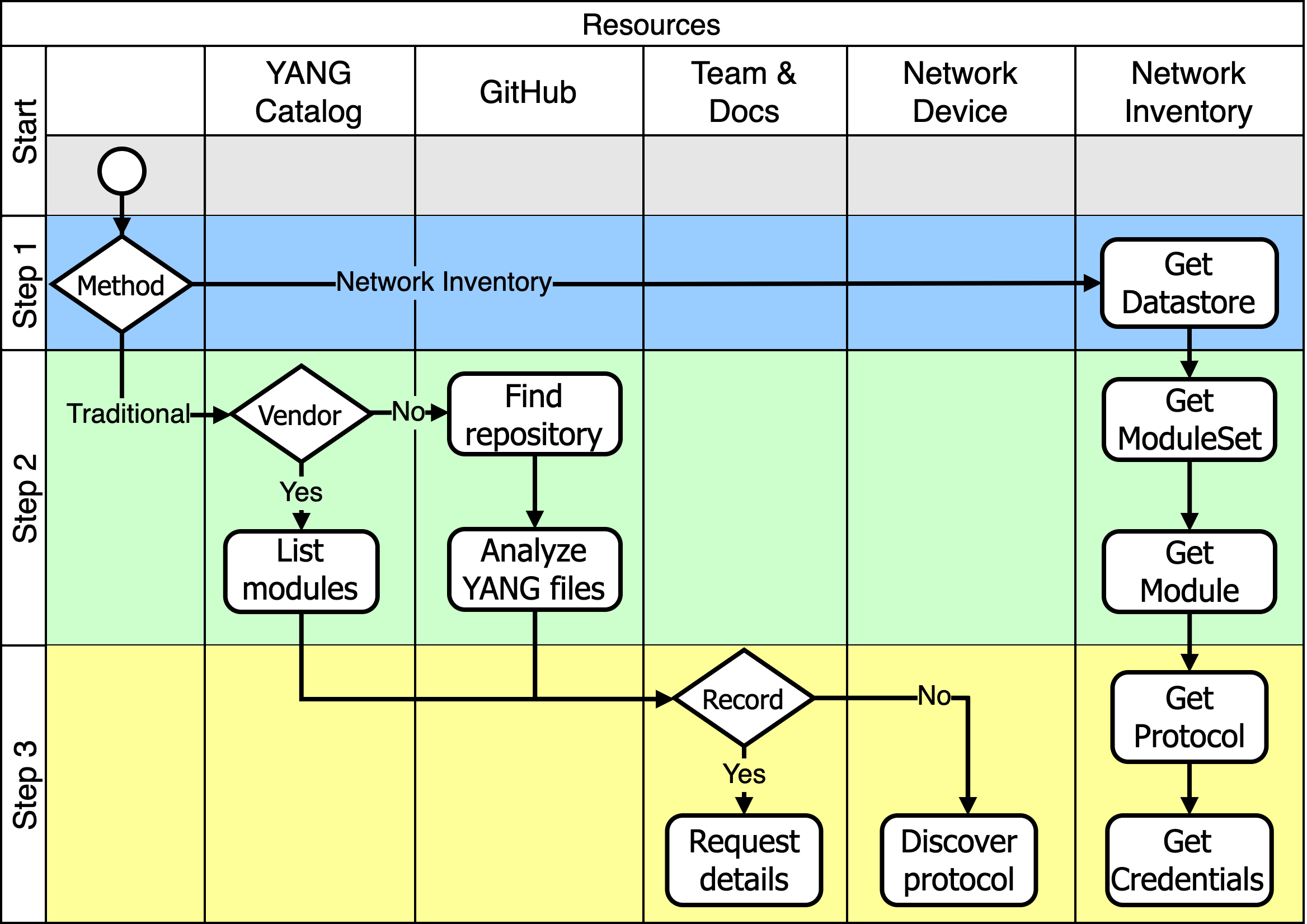}
\caption{Workflow diagram of different traditional and  inventory-based methods.}
\label{fig:workflow}
\end{figure}

Suppose a NetOps engineer wants to monitor the interfaces of different devices that comprise a multi-vendor network. The goal of the engineer is to develop a monitoring application that, for instance, could compute network interface KPIs like packet loss. To this end, the engineer must discover what YANG data pertaining to interface management can be obtained, and also, discover how these data can be collected through protocols like NETCONF or gNMI. However, the engineer is not familiar with the monitored network, and only knows about the models of the network devices and, possibly, IP address and credentials information for SSH access to the devices. Fig. \ref{fig:workflow} depicts the different workflows that the engineer would take depending on the used method and network vendor. These workflows are divided into three common steps that represent questions that the engineer seeks to answer as described in the following.

\textbf{Step 1: Identifying available datastores}. For those network devices that support NMDA, finding the available datastores is of high importance. For example, the intended status of an interface could be disabled whereas the actual status could be enabled. The discovery of the different datastores implemented by a given device is not possible through the traditional approach because resources like YANG Catalog do not track NMDA datastores. In this sense, the proposed Network Inventory is the only resource that can provide NMDA information, which can be easily obtained by listing the \textit{Datastore} entities linked to the target \textit{Platform} entity. However, our experiments show that NMDA support is at an early stage as Cisco is the only vendor that supports this new architecture.

\textbf{Step 2: Identifying implemented YANG modules for interface management}. This is the most complex step since it requires finding all YANG modules related to interface management such as \textit{openconfig-interfaces}. In the traditional method, YANG Catalog can provide the list of modules implemented per network device as in the case of Cisco or Huawei. However, vendors like Nokia are not registered in YANG Catalog, which requires the engineer to find a public repository where the YANG files for the specific device model are stored. Then, these files need to be fetched and analyzed in order to find those modules of interest. On the other hand, the Network Inventory can provide the YANG modules of interest in just a few queries. First, the \textit{ModuleSet} entity for the target \textit{Platform} is retrieved. Next, for such \textit{ModuleSet} entity, the list of \textit{Module} entities related to interface management are collected.

\textbf{Step 3: Collecting MDT protocol information}. Once the YANG modules of interest have been identified, the goal of this last step is to understand how the YANG data can be collected through MDT. To this end, engineers need to discover, for each device, the features of their specific MDT protocol implementation, namely, IP address, port, credentials, and protocol capabilities. In this regard, the traditional method typically entails looking for team partners that may know about this information. However, there are many scenarios in which there are no records of this information. For those cases, the engineer must interact directly with the device, either using YANG explorers or using custom applications. When using the Network Inventory, this step is greatly simplified, as engineers can discover the protocol implementation details by retrieving the \textit{Protocol} entities of a particular \textit{Platform}. The results from this experiment have proven, for example, that protocol information will help in determining how YANG data can be selected for streaming. In the case of Huawei, its implementation of NETCONF includes support for XPath filtering, whereas Cisco and Nokia do not support this feature, hence, they would only allow for the native subtree filtering. 

\section{Conclusions and Future Work}

This paper has introduced the Network Inventory: a centralized service to facilitate the application of MDT for network monitoring. By building on the ETSI CIM standard, context information related to MDT can be easily integrated as a property graph.

To integrate such information, new NGSI-LD information models have been defined for different isolated domains of information, namely, network platforms and module catalogs. The resulting models prove that these domains can be linked by having the YANG module and submodule entity types in common.

To support these NGSI-LD information models within the Network Inventory, a reference architecture has been designed. This architecture automates the collection of context information from each domain by making use of the standard NGSI-LD API. As a result, the proposed Network Inventory becomes a one-stop-shop for NetOps engineers where to discover relevant context information related to MDT such as the supported YANG modules or the implementations of network management protocols.

The validation of the Network Inventory has been carried out through the design of a use case and the development of a prototype. The experiments conducted have demonstrated that our proposed inventory-based method is vendor-agnostic and centralizes all relevant information, unlike the traditional method which requires manual interactions that may vary depending on the vendor.  

Furthermore, our inventory-based method not only simplifies the discovery of MDT information compared to the traditional method, but also allows for integrating new context information from other isolated domains. In this sense, for future works we intend to incorporate new domains of information such as network topology. From a traditional perspective, this domain contains context information that represents physical connections between network devices. With the advent of the NFV paradigm, this context information conveys connections between virtualized network functions.

The approach of this work has focused on managing YANG data models to facilitate network monitoring by means of MDT. YANG is a data modeling language that enables differentiating configuration from operational data. Therefore, the proposed Network Inventory could be leveraged to determine configurable information in YANG-enabled devices.

Additionally, the ETSI CIM standard opens a new door to applying techniques of the Semantic Web on context information. In this sense, we plan to investigate on semantic reasoners, which could be leveraged to analyze the impact of a deprecated YANG module in a network infrastructure.

%
%
\section*{Acknowledgment}
This work has been partly funded by the European Union's Horizon 2020 research and innovation program under Grant Agreement No. 871428 (5G-CLARITY project), the Spanish Ministry of Economy and Competitiveness, and the Spanish Ministry of Science and Innovation in the context of ECTICS project under Grant PID2019-105257RB-C21 and Go2Edge under Grant RED2018-102585-T

%
%
\bibliographystyle{ieeetr}
\bibliography{references}

\begin{thebibliography}{10}

\bibitem{Claise_book}
B.~Claise, J.~Clarke, and J.~Lindblad, {\em Network Programmability with YANG: The Structure of Network Automation with YANG, NETCONF, RESTCONF, and gNMI}.
\newblock Addison-Wesley Professional, 2019.

\bibitem{mdt_bgp}
R.~A.~K. Fezeu and Z.-L. Zhang, ``Anomalous model-driven-telemetry network-stream bgp detection,'' in {\em 2020 IEEE 28th International Conference on Network Protocols (ICNP)}, pp.~1--6, 2020.

\bibitem{Pastor2021}
A.~Pastor, D.~R. López, J.~Ordonez-Lucena, S.~Fernández, and J.~Folgueira, ``{A Model-based Approach to Multi-domain Monitoring Data Aggregation},'' {\em Journal of ICT Standardization}, vol.~9, pp.~291--310, 2021.

\bibitem{yang_stats}
``Yang modules stats - yangcatalog.org.'' \url{https://yangcatalog.org/private-page}.
\newblock Accessed on: 2022-08-30.

\bibitem{yang_models}
``Yang models.'' \url{https://github.com/YangModels/yang}.
\newblock Accessed on: 2022-05-23.

\bibitem{yang_catalog}
``Yang catalog.'' \url{https://yangcatalog.org/}.
\newblock Accessed on: 2022-05-23.

\bibitem{yang_suite}
``Cisco yang suite.'' \url{https://developer.cisco.com/yangsuite/}.
\newblock Accessed on: 2022-05-23.

\bibitem{cim_006}
ETSI, ``{ETSI GS CIM 006 - V1.1.1 - Context Information Management (CIM); Information Model (MOD0)},'' July 2019.

\bibitem{cim_009}
ETSI, ``{ETSI GS CIM 009 - V1.5.1 - Context Information Management (CIM); NGSI-LD API},'' November 2021.

\bibitem{Lopez-Morales2020}
J.~A. L{\'{o}}pez-Morales, J.~A. Mart{\'{i}}nez, and A.~F. Skarmeta, ``{Digital transformation of agriculture through the use of an interoperable platform},'' {\em Sensors (Switzerland)}, vol.~20, no.~4, pp.~1--20, 2020.

\bibitem{Jeong2020}
S.~Jeong, S.~Kim, and J.~Kim, ``{City data hub: Implementation of standard-based smart city data platform for interoperability},'' {\em Sensors (Switzerland)}, vol.~20, no.~23, pp.~1--20, 2020.

\bibitem{rfc8525}
A.~Bierman, M.~Björklund, J.~Schönwälder, K.~Watsen, and R.~Wilton, ``{YANG Library}.'' RFC 8525, Mar. 2019.

\bibitem{clacla-netmod-model-catalog-03}
J.~Clarke and B.~Claise, ``{YANG module for yangcatalog.org},'' Internet-Draft draft-clacla-netmod-model-catalog-03, Internet Engineering Task Force, Apr. 2018.
\newblock Work in Progress.

\bibitem{network-inventory}
``Network inventory prototype.'' \url{https://github.com/giros-dit/network-inventory}.
\newblock Accessed on: 2022-05-23.

\bibitem{scorpio}
``Scorpio ngsi-ld broker.'' \url{https://github.com/ScorpioBroker/ScorpioBroker}.
\newblock Accessed on: 2022-05-23.

\end{thebibliography}

%
%
\section*{Biographies}
\textbf{Ignacio D. Martinez-Casanueva} is a PhD candidate at Universidad Politecnica de Madrid, Spain.

\textbf{Daniel Gonzalez-Sanchez} is a PhD candidate and Teaching Assistant at Universidad Politecnica de Madrid, Spain.

\textbf{Luis Bellido} is an Associate Professor at Universidad Politecnica de Madrid, Spain.

\textbf{David Fernandez} is an Associate Professor at Universidad Politecnica de Madrid, Spain.

\textbf{Diego R. Lopez} is a Senior Technology Expert in Telefonica’s GCTIO team.

\end{document}